\magnification=\magstep1
\rightline{TAUP 2077/93}
\rightline{IASSNS-HEP-92/74}
\rightline{7 February, 1997}
\bigskip
\centerline{\bf Schwinger Algebra}
\centerline{\bf for}
\centerline{\bf Quaternionic Quantum Mechanics}
\bigskip
\centerline{L.P. Horwitz\footnote{\dag}{Also at Department
of Physics, Bar Ilan University, Ramat Gan Israel. }}
\centerline{ School of
Physics and Astronomy, Raymond and Beverly Sackler Faculty of Exact
Sciences}
\centerline{ Tel Aviv University, Ramat Aviv, 69978 Israel}

\bigskip
\noindent
PACS

\noindent {\it Abstract:\/} It is shown that the measurement algebra
of Schwinger, a characterization of the properties of Pauli
measurements of the first and second kinds, forming the foundation of
his formulation of quantum mechanics over the complex field, has a
quaternionic generalization. In this quaternionic measurement algebra
some of the notions of quaternionic quantum mechanics are clarified.
The conditions imposed on the form of the corresponding quantum field
theory are studied, and the quantum fields are constructed. It is
shown that the resulting quantum fields coincide with the fermion or
boson annihilation-creation operators obtained by Razon and Horwitz in
the limit in which the number of particles in physical states $N
\rightarrow \infty$.

\eject
\bigskip
\noindent
 {\bf 1. Introduction}
\smallskip
\par Schwinger$^1$ has given an interpretation of the structure of the quantum
theory,
 which is
 both deep and pedagogically useful, in terms of the algebraic properties of
filters.
These filters are considered to represent the selection of subensembles from a
``beam''
of independent, but identical, quantum systems, corresponding to the quantum
ensemble
containing the properties of a single system. The algebra of these filters is
represented
in a vector space that is the Hilbert space of the quantum theory. It is assumed
that
the coefficients which arise when a sequence of measurements of non-compatible
observables is made are elements of the complex number field {\bf C}.  In this
work,
I generalize this structure to admit elements of the quaternionic
algebra {\bf H}, where, for $q \in {\bf H}$,  the real algebra of
quaternions (with involutory automorphism $q^*$),
$$\eqalign{ q&= \sum_0^3 \lambda_i e_i \ (\lambda \  {\rm real});\cr
 e_i e_j &= e_k (ijk \ {\rm cyclic}),\qquad e_i^* = -e_i,\ \ i \neq 0,\cr
e_ie_j &= -e_je_i \ \ (i\neq j \neq 0),\cr
 e_0 =1,\ & {\rm and }\ \  e_i^2 =-1,\cr}
$$
 and show that the resulting vector space is the Hilbert module corresponding to
the
quaternionic quantum theory$^{2,3,4}$.
\par Schwinger$^1$ proceeds to construct quantum field theory by considering
each
system as a composite of a set of identical subsystems.  The corresponding set
of
measurement filters are then constructed as a (direct) sum of filters sensitive
to each of the
subsystems.  Since the subsystem filters commute, the algebra may be closed as a
commutator algebra, and the factorized solutions for the representations of
these
operators result in the definition of the quantum fields.  In the case of
quantum
fields on the usual complex Hilbert space, the numerical coefficients
corresponding
to the transformation functions (transition amplitudes) factor out of the
non-linear
algebraic expressions, but in the non-commutative case that I consider in this
work,
the requirement of Bose-Einstein or Fermi-Dirac symmetry imposes a structure on
the
numerical coefficients according to which they must act distributively, as
derivations.
With this structure, I find that the quantum fields defined by Schwinger's
method
correspond to the annihilation-creation operators of the Fock space constructed
by Horwitz and Razon$^5$, in the limit in which the physical states of the
theory
contain occupation number $N \rightarrow \infty$.  It is only in this case that
the commutation (anticommutation) relations are not deformed$^6$, so that
bilinears
in these operators may generate closed Lie algebras.
\par In the following, I review Schwinger's construction briefly for the complex
(Abelian) case, and in Section 2, construct the corresponding quaternionic
measurement
algebra and its representation in terms of the quaternionic Hilbert module.  In
Section 3,
the measurement algebra for the non-commutative quaternionic many-body system of
identical particles is constructed, and in Section 4, I construct its
representation in
terms of quantum fields.  Some conclusions and discussion are given in Section
5.
\par Schwinger considers a system defined by a set of observables $A, B, C,
\dots $, with possible real values $a', b', c',  \dots $.  Let us
assume the existence of filters$^1$, symbolized by $\{M(a')\}$, and a {\it
beam} of systems which have these properties; the filter $M(a')$ selects all
systems
with the property $A=a'$ from the beam and passes them with this
observable having the exact value $a'$ immediately after the
measurement. This
operation corresponds to a Pauli measurement of the
first kind (the spectra $\{a'\}$ may be discrete or continuous;
discussion of the latter case is technically somewhat more
complicated, and we do not treat it explicitly here).   It then
follows that
$$ M(a'')M(a') = \delta (a',a'') M(a'), \eqno(1.1)$$
provided that the filters act exclusively on the possible values
$\{a'\}$. In addition to this orthogonality property, the
interpretation of addition as {\it or} is manifest in the completeness
statement,
$$ \sum_{a'} M(a') = I, \eqno(1.2),$$
where $I$ is the measurement which selects and transmits every system
in the beam, without alteration.  To realize this contruction, one may
think of storing each outcome of the action of $M(a')$ in a storage
ring; repeating the process for each $a'$ results in a stored beam
identical to the original one.
\par Schwinger defines the Pauli measurement of the second kind, symbolized by
$M(a',b')$; it corresponds to a filter which selects systems with value $b'$ of
the property $B$, and passes them through the filter with property $A$ having
definitely the value $a'$ (e.g., systematic change of polarization).
\par   If the
properties $A,\ B$ are not compatible, a measurement of the first kind
selecting systems with value $b'$ of property $B$ leaves the set of systems with
a distribution of possible values $a'$ of $A$ (eventually to be determined by
the
physical nature of the system).  Following the $B$ measurement, an  $A$
measurement leaves the system with a definite value $a'$ of $A$.  This operation
is represented by
$$ M(a') M(b') = \langle a' \vert b' \rangle M(a',b'), \eqno(1.3)$$
since the selected systems have definite value $b'$ of $B$, and are left with
a definite value $a'$ of $A$; the uncertainty in finding $a'$ in a system with
the value $b'$ of $B$ reflects the microscopic indeterminacy of the quantum
theory, and is represented by
the coefficient $\langle a' \vert b' \rangle$. The closure of the
algebra in the form
$$ M(a',b') M(c',d') = \langle b' \vert c' \rangle M(a',d')
\eqno(1.4)$$
clearly depends on the assumption of Abelian $\{\langle b'\vert c'
\rangle\} $.  Using $(1.2)$,
$$ \eqalign{M(a') \delta(a',a'') &= \sum_{b'} M(a')M(b')M(a'')\cr
                                 &= \sum_{b'}  \langle a' \vert
b' \rangle \langle b' \vert a'' \rangle M(a',a''), \cr} \eqno(1.5) $$
where we have used $M(a',a') \equiv M(a')$.  We therefore have the
completeness relation
$$ \delta(a',a'') = \sum_{b'} \langle a' \vert b' \rangle \langle b'
\vert a'' \rangle. \eqno(1.6)$$
\par Schwinger$^1$ defines a linear map from the set $M(a',b')$ to the
number system of the coefficients $\langle a' \vert b' \rangle$,
$$ {\rm tr} M(a',b') = \langle b' \vert a' \rangle; \eqno(1.7) $$
with the definition of an adjoint action (reversal of the second kind
filtering process),
$$ M(a',b') ^\dagger
= M(b',a'), \eqno(1.8) $$
so that
$$\eqalign{ {\rm tr} M(b',a') = {\rm tr} \bigl(M(a',b')^\dagger \bigr)
&= \bigl({\rm tr} M(a',b')
\bigr)^\dagger = \langle b' \vert a' \rangle^\dagger  \cr
&= \langle a' \vert b' \rangle; \cr} \eqno(1.9) $$
identifying this action on the number system with its involutory
automorphism, $(1.6)$, for $a'=a''$, becomes, if we assume that we are
dealing with a normed algebra,
$$ 1= \sum_{b'} \vert \langle b' \vert a' \rangle
\vert^2. \eqno(1.10)$$
It is then assumed that the $\{\langle b' \vert a' \rangle \}$ are in
${\bf C}$.
\par Schwinger then asks for a representation of this algebra in
terms of a vector space.  The factorized definition
$$ M(a',b') = \vert a' \rangle \langle b'\vert, \eqno(1.11)$$
with the rule $\langle b' \vert \cdot \vert c' \rangle = \langle b'
\vert c' \rangle$, constituting a scalar product in this vector space,
satisfies all the structural requirements imposed by the algebra.  The
completion of this vector space is the usual quantum mechanical
Hilbert space over the complex field.
\par  We do not wish to reproduce here the large set of consequences of
the structure outlined above, but only to emphasize the basic
properties of the algebra.  As we shall see, the assumption that the
$\{\langle a' \vert b' \rangle\} \in {\bf C} $ is not necessary; the
quaternion skew-field ${\bf H}$ provides an equally workable choice,
and includes a higher level of intrinsic structure$^{2,3,4}$.
\par We now describe briefly the structure of the many-body theory, in
the context of the complex number field.  To deal with the many-body
theory, Schwinger considers the situation in which each system
(element of the beam) consists of a set of identical subsystems
carrying the same set of properties $A,B,C, \dots$.  Filters for such
a beam can be constructed in a symmetrical way as
$$ {\bf M}(a',b') = \sum_{i=1}^N M(a_i', b_i'), \eqno(1.12)$$
where each $M(a_i', b_i') $ is a filter of the second kind acting on
the subsystem labelled by $i$. The entire system is passed if the
requirements on any subsystem (that $B$ of $i$ have the value $b'$;
it is then transformed to $a'$ of $A$) are met.  The beam resulting
from the sum contains (relatively) as many systems as there were
in the incoming beam which have a subsystem with the value $b'$ of $B$
(in correspondence with the action of an annihilation operator on a
state in the Fock space); in the final state, these subsystems have
value $a'$ of $A$.
\par The action of $M(a_i', b_i')$ and $M(c_j', d_j')$, for $i \neq
j$, commute, since the subsystems are considered to be dynamically
independent.  The algebra of the many-body measurement symbols
$(1.12)$ then closes as a {\it commutator} algebra in the form
$$\eqalign{ {\bf M}(a',b') {\bf M}(c',d') &- {\bf M}(c',d') {\bf
M}(a',b') \cr
&= \langle b'\vert c' \rangle {\bf M}(a',d') - \langle d' \vert a'
\rangle {\bf M}(c',b'), \cr} \eqno(1.13)$$
where the isomorphic structure of all the subsystems implies that
$\langle a_i' \vert b_i' \rangle \equiv \langle a' \vert b' \rangle$
are all equal.  The factorization
$$ {\bf M}(a',b') = \psi(a')^\dagger \psi(b'), \eqno(1.14) $$
where
$$[\psi(a'), \psi(b')^\dagger]_\mp = \langle a' \vert b' \rangle
\eqno(1.15)$$
provides an effective solution for the algebra $(1.13)$.  With this,
Schwinger is able to identify the quantum fields.
\par In the succeeding sections, I discuss a generalization of this
structure to the quaternionic skew-field.
\par This paper is dedicated to
the memory of my former friend and colleague, Larry Biedenharn, with
whom there were many years of collaboration in the study of
hypercomplex quantum theories, among other subjects; it also intended as
 a tribute to
his courageous wife, Sarah, who helped to make this collaboration a
warm friendship.
\bigskip
\noindent{\bf 2. Measurement algebra of the quaternionic quantum theory}
\smallskip
\par To construct an analog of the Schwinger algebra which results in
the Hilbert module of quaternionic quantum mechanics, we assume that
there is a non-trivial (non-Abelian) degree of freedom$^2$ associated with
the algebraic orientation of the number field associated with the
final state of a system emerging from the filters corresponding to
ideal measurements.  Such a filter is empowered with the possibility
of introducing an algebraic automorphism on the number system.  I
 represent this as $M(a';q;b')$, a generalized Pauli measurement of
the second kind, where, in the quaternion case in which we are
interested, $q \in {\bf H}$.  It arises in the action
$$ M(b') M(a') = M(b'; \langle b' \vert a' \rangle;a'), \eqno(2.1) $$
where the number $\langle b' \vert a' \rangle$, a measure of the
likelihood of finding $b'$ among a set of systems with $A$ having
value $a'$, contains as well an essential non-Abelian phase.
 We assume, moreover, the linearity properties
$$ \eqalign{\sum_i M(a';q_i;b') &= M(a'; \sum_i q_i ; b')\cr
                 M(a'; \lambda q ;b') &= \lambda M(a';q;b'), \cr}
\eqno(2.2)$$
for $\lambda$ real.
\par The filter $M(b';q;a')$ selects a system for which the observable
$A$ has the value $a'$ and passes this system with observable  $B$
having, immediately after the measurement, value $b'$ generating, at
the same time, an automorphism induced by $q$ on the structure of the
number system ( in this case, quaternionic) to be used for  the
description of  the {\it state}  of a
system for which the observable $B$ has the value $b'$ (this
observable could, of course, as a special case, be $A$).  In
introducing this notion, we  have explicitly recognized that an
observable carries an intrinsic degeneracy.  An analogous property
exists for the complex number system in the usual formulation of the
quantum theory; since it is commutative, it may be separated, as a phase,
multiplicatively from the measurement symbols, along with the real
numerical size of the coefficient (the local phases emerging at a
later stage  of
development of the theory result in the $U(1)$ bundle characterizing
electromagnetic interactions). In Eq.$(2.1)$, the coefficient carries
the same meaning implied by Schwinger$^1$; a system with $A$ having
the value $a'$ has $B$ with a distribution of values $b'$, represented
by $\langle b' \vert a' \rangle$, to be
eventually determined by the dynamical properties of the system.  It
carries, moreover, the additional information
associated with the change of orientation of the number systems used
for the $A$ and $B$ descriptions, in particular,  relating
the spectral values $a'$ and $b'$ {\it locally}.  Such structures, with
noncommutative degrees of freedom, associated with the manifold on which
functions are to be defined (resulting, as we shall see, in sets of
noncommutative functions) is sometimes referred to as noncommutative
geometry$^{7,8}$.
\par   The algebra  of measurements is invariant
under such automorphisms, which therefore constitute what might be
considered a symmetry of the
``apparatus'', in the sense of Piron$^9$.
\par The generalized Pauli measurements of the second kind satisfy
$$M(a';q;b')M(c';p;d') = M(a'; q\langle b'\vert c'\rangle p;d').\eqno(2.3)$$
The ``completeness sum''
$$ \sum_{a'} M(a';e_i;a') \equiv E_A(e_i) \eqno(2.4)$$
satisfies an algebra isomorphic (it may be constructed to be
star-isomorphic, as we show below) to ${\bf H}$, and it is called$^{3,4}$
an element
of the  {\it left quaternion algebra} associated with the spectrum of
$A$ ($M(a'; e_0; a')\equiv M(a')$, so according to $(1.2)$, $E_A(e_0)
= I$).
\par With Eqs.$(1.2)$ and $ (2.1)$, we see that, as for the complex theory,
$$ \eqalign{\sum_{b'} M(a')M(b')M(a'') &= M(a')M(a'')=
\delta(a',a'') M(a') \cr
&= \sum_{b'} M(a'; \langle a' \vert b' \rangle \langle b' \vert a''
\rangle ; a''),\cr} $$
so that
$$ \sum_{b'} \langle a' \vert b' \rangle \langle b' \vert a'' \rangle =
\langle a' \vert a'' \rangle =\delta (a',a''). \eqno(2.5)$$
More generally,
$$ \eqalign{M(a')M(c')&= \sum_{b'} M(a')M(b')M(c') \cr
                       &=\sum_{b'} M(a';\langle a'\vert b' \rangle
\langle b' \vert c' \rangle ; c') , \cr}$$
and hence
$$ \sum_{b'} \langle a' \vert b' \rangle \langle b' \vert c' \rangle =
\langle a' \vert c' \rangle. \eqno (2.6)$$
As for the complex theory, the quantitites $\langle a' \vert b'
\rangle$ therefore act as transformation functions.  To see their
action on the change of description in the measurement symbols,
consider the relation
$$ M(a'; q; b') = \sum_{c'} M(a';q;b') M(c') = \sum_{c'} M(a';
q\langle b'  \vert c'
\rangle; c'). \eqno(2.7)$$
This transformation law can be written in a form more analogous to
that of the complex theory by using elements of the left algebra:
$$\eqalign{\sum_{c'}M(a';q;c')E_C(\langle b' \vert c' \rangle) &=
\sum_{c',c''} M(a';q;c')M(c'';\langle b' \vert c' \rangle;c'')\cr
&= \sum_{c'} M(a'; q \langle b' \vert c' \rangle ;c')\cr
&=  M(a';q; b'), \cr} \eqno(2.8)$$
according to Eq. $(2.7)$.  Similarly, one finds
$$ M(a';q;b') = \sum_{c'} E_C (\langle c'\vert a' \rangle) M(c';q;b').
\eqno(2.9)$$
\par Following Schwinger$^1$, we define the conjugation
$^\dagger$ for which $M(a';e_0;b')\equiv M(a',b')$ satisfies
 $$M(a',b')^\dagger = M(b',a'), \eqno(2.10) $$
 and
$$(M(a',b')M(c',d'))^\dagger = M(d',c')M(b',a').\eqno(2.11)$$
\par If we supplement the definition $(2.2)$  by defining the action
of $^\dagger$ on the number system entering the generalized Pauli
measurement of the second kind through
$$ M(a'; q; b')^\dagger = M(b';q^*;a'), \eqno(2.12)$$
where $^*$ is the conjugation induced by $^\dagger$ on the number
system ${\bf H}$,
it then follows from $(2.11)$ that
$\langle b' \vert c' \rangle ^* = \langle c' \vert b' \rangle$.  Eqs.
$(2.4)$ and $(2.12)$ imply that $E_A(p)^\dagger = E_A(p^*)$. If
we identify the action of this conjugation with an
involutory automorphism isomorphic to that of ${\bf H}$,
 the left algebra is then star- isomorphic to
${\bf H}$.  Eq.$(2.6)$ then becomes, for $c' = a'$,
$$ \sum_{b'} \vert \langle a' \vert b' \rangle \vert^2 = 1, \eqno(2.13)$$
providing a formal probability interpretation, as for the complex
quantum theory, for the absolute square of the amplitude $\langle a'
\vert b' \rangle $. \footnote{$\natural$}{The use of Clifford algebras
of higher
order than the $C_2$ quaternions requires some modification (e.g.,
introduction of the trace) to achieve
an analogous result, since they are not normed division algebras. This more
general case will be discussed elsewhere.}
\par In the quaternionic
theory, a real linear map corresponding to $(1.7)$ from the
measurement
 algebra to ${\bf H}$
can be defined, but has properties that are somewhat more complicated.
\par Let us define a map from the measurement algebra to ${\bf H}$ by
$$ {\rm tr}_A M(a'; q; b') = q \langle b' \vert a' \rangle \eqno(2.14)$$
and
$$ {\rm tr}_B M(a'; q; b') = \langle b' \vert a' \rangle q. \eqno(2.15)$$
These are not, in general, equal (in the complex case, they are equal, and
independent of the basis used for the evaluation).
It follows, however, from $(2.7)$, that
$$ {\rm tr}_C M(a';q;b') = \sum_{c'} \langle c' \vert a' \rangle q \langle b'
\vert c' \rangle, \eqno(2.16)$$
\par  Taking the real
part of $(2.14)$ or $(2.15)$, equivalent to carrying out an additional
trace
 over the matrix
representation of ${\bf H}$, removes this dependence;  the resulting
operation, called the {\it total trace}$^4$, admits cycling the transformation
coefficients.  Adler$^{4,10,11}$ has shown that this  total trace
operation plays an essential role in constructing a dynamical theory,
and permits a wide generalization of the structure (in its application
to quantum field theory, a graded total trace is used).
\par Elements of the measurement algebra called {\it operators} can be
constructed in analogy with Schwinger's definition.  A general operator
is taken to have the form
$$ {\cal O} = \sum_{a',b'} M(a'; \langle a'\vert {\cal O} \vert b'
\rangle; b'). \eqno(2.17)$$
In another basis,
$$ {\cal O} = \sum_{a',c'} M(a'; \langle a' \vert {\cal O} \vert c'
\rangle ; c'). \eqno(2.18)$$
Using $(2.8)$ to rewrite $(2.18)$, i.e.,
$$ {\cal O} = \sum_{a', b',c'} M(a'; \langle a' \vert {\cal O} \vert
c'\rangle \langle c' \vert b' \rangle  ; b'), $$
we see that
$$ \langle a' \vert {\cal O} \vert b' \rangle = \sum_{c'} \langle a'
\vert {\cal O} \vert c' \rangle \langle c' \vert b' \rangle;
\eqno(2.19)$$
the quantities $\{\langle a' \vert b'\rangle\}$ therefore act as
transformation functions on the quaternionic coefficients
characterizing the operators as well.  That the representation
$(2.17)$ induces a homomorphic structure on these coefficients can be
seen by examining the representation of a product of two operators
${\cal O}_1$ and ${\cal O}_2$:
$$ \eqalign{{\cal O}_1 {\cal O}_2 &= \sum_{a',b',a'',b''} M(a'; \langle a' \vert
{\cal O}_1 \vert b' \rangle ; b') M(a''; \langle a'' \vert {\cal
O}_2\vert b''\rangle; b'')\cr
&= \sum_{a',b',a'',b''} M(a'; \langle a' \vert {\cal O}_1 \vert
b'\rangle \langle b' \vert a'' \rangle \langle a'' \vert {\cal O}_2
\vert b''\rangle; b''); \cr} \eqno(2.20)$$
With the help of $(2.19)$, one sees that $(2.20)$ can be written as
$$ {\cal O}_1{\cal O}_2 = \sum_{a', b'} M(a'; \sum_{a''}\langle a' \vert
{\cal O}_1 \vert a'' \rangle \langle a'' {\cal O}_2 \vert b'\rangle;
b'). \eqno(2.21)$$
 \par Returning to the general definition $(2.17)$, the transformation
theory we have described permits us to rewrite it entirely in terms of
the spectrum of a single observable $C$, i.e.,
$$ {\cal O} = \sum_{c',c''} M(c'; \langle c' \vert {\cal O} \vert c''\rangle
; c''); \eqno(2.22)$$
if ${\cal O}^\dagger = {\cal O}$, i.e., a self-conjugate operator of the
measurement algebra, it follows that
 $$ \langle c' \vert {\cal O} \vert c'' \rangle ^* = \langle c''\vert
{\cal O} \vert c' \rangle. \eqno(2.23)$$
It follows from the spectral theory of a quaternionic self-adjoint
operator$^{3,4}$, realized in this case by the matrix $\{\langle c'\vert
{\cal O} \vert c'' \rangle \}$, that there exists a unitary
transformation to a representation in which the matrix is diagonal.
Expressing the operators $M(c'; q; c'')$ in terms of the spectral
family of another observable, say, $A$, one obtains
$$ {\cal O} = \sum_{a', a'', c', c''} M(a'; \langle a' \vert c'
\rangle \langle c' \vert {\cal O} \vert c'' \rangle \langle c'' \vert
a''\rangle ; a''). \eqno(2.24)$$
The transformation coefficients act on the representative of ${\cal
O}$ as a unitary transformation; with an appropriate choice of $A$,
the resulting matrix becomes diagonal, and the operator then takes on
the form
$$ {\cal O} = \sum_{a'} {\cal O}_{a'} M(a'), \eqno(2.25)$$
where $\{ {\cal O}_{a'} \}$ are real.
The trace,
$$ {\rm tr} {\cal O} = \sum_{a'} {\cal O}_{a'}, $$
is then independent of the complete set used for the evaluation.
\par In the case of a general operator, $(2.22)$ is clearly not, in general,
diagonalizable to a real spectrum, and the trace is representation
dependent.  For example,
$$ {\rm tr}_A {\cal O} = \sum_{a',c',c''} \langle a' \vert c' \rangle
\langle c' \vert {\cal O} \vert c'' \rangle \langle c'' \vert a'
\rangle, $$
which is not equal to
$$ {\rm tr}_C {\cal O} = \sum_{c'} \langle c' \vert {\cal O} \vert c'
\rangle; $$
for the {\it total} trace, the transformation coefficients can be cycled,
and the result becomes independent of the basis.
\par Again, following Schwinger$^1$, a natural representation for the
quaternionic measurement algebra is
  $$ M(a';q;b') = \vert a'\rangle  q \langle b' \vert,\eqno(2.26) $$
with the rule
$$ \langle a'\vert \cdot \vert b' \rangle = \langle a'\vert b'\rangle,
\eqno(2.27)$$
i.e., the formation of the scalar product.  We then see that if
$$M(a',b') \vert c' \rangle = \vert a' \rangle \langle b' \vert c' \rangle,$$
it follows that
$$ M(a';q;b') \vert c' \rangle = \vert a' \rangle \bigl( q\langle b'
\vert c' \rangle q^{-1} \bigr) q, \eqno(2.28)$$
inducing an automorphism on the algebra of transformation
coefficients.
\par This representation clearly reproduces all of the definitions and
results stated above, and corresponds to the structure of a
quaternionic Hilbert space.$^{2,3,4}$
\bigskip
\noindent
{\bf 3. Identical particles and quantum fields}
\smallskip
\par We now turn to the formulation of the many-body problem and the
construction of quantum fields, for the case that the transformation
functions belong to ${\bf H}$, according to the algebraic method of
Schwinger$^1$.
\par Again,we think of every element (system) in the beam, on which filtering
is performed, as constructed of a number $N$ of identical subsystems, each
with an identical set of properties $A,B,C,\dots$. Enumerating the
possible value these properties can acquire as $a_i', b_i', \dots$ for
$i=1,2,\dots N$ labelling the subsystem, the subsystem filters are
denoted as $M(a_i'), M(b_i'), \dots$.  A symmetrical filter for a
system of such subsystems can then be constructed as the (direct) sum
$$ \sum_i M(a_i') = {\bf M}(a'). \eqno(3.1)$$
Recall that the interpretation of the sum of measurement symbols over
possible values $a'$ of $A$, as in Eq.$(1.2)$, refers to the retention
of elements of the beam in some storage facility.  We may therefore
consider the outcome of the action of $(3.1)$ as the collection of
$N$-body systems for which the $i^{th}$ subsystem has the value $a_i'$
of $A$.
\par I shall assume that the value of a variable occurring on one
subsystem is independent of the values occurring on other subsystems,
hence the measurement symbols $\{M(a_i';q;b_i')\}$ commute for $i \neq
j$.  The composite measurements
$$ {\bf M}(a';q;b') = \sum_i M(a_i';q;b_i') $$
should then form a closed algebra under commutation, as for the
Abelian case.  Some care must be taken, however, in the formation of numerical
coefficients as the result of a product of two measurement symbols
corresponding to the same subsystem.  As pointed out in the
one-particle case for the interpretation of $M(b';q;a')$, after
selection of subsystems with value $a_i'$ of $A$, the element of the
beam (the whole system) is passed with the value $b_i'$ of $B$, and an
automorphism is generated on the numbers used to describe the state of
the subsystem.  As we shall see, the requirement that the system
retain its symmetry after the measurement requires that this
automorphism be induced by this operation on all of the subsystems in
turn.  The phase disturbance induced on one subsystem propagates, by
means of the manifest symmetry of the system, to all of the subsystems
in the sense of a derivation (corresponding to the additive effect of
an infinitesimal transformation on a product, i.e., the Leibniz rule).
\par Consider, for example, the two body case, where we represent the
physical state as (according to Bose-Einstein or Fermi-Dirac symmetry)
$$ \Psi(c',d') = {1 \over \sqrt{2}} \bigl( \vert c_1' \rangle \otimes
\vert d_2' \rangle \pm \vert d_1' \rangle \otimes \vert c_2' \rangle
\bigr). \eqno(3.1) $$
The indices refer to the particle names, and $c',\ d'$ specify the
states of the subsystems $1,\ 2$.  The direct action of ${\bf
M}(a',b')$ on $\Psi(c',d')$ is given (naively) according to the
relations obtained in the one-particle theory, by
$$ \eqalign{\bigl(M(a_1', b_1') + M(a_2', b_2')\bigr) {1 \over \sqrt{2}}
 \bigl( \vert c_1' \rangle \otimes
\vert d_2' \rangle &\pm \vert d_1' \rangle \otimes \vert c_2' \rangle
\bigr)\cr  &={1 \over \sqrt{2}} \bigl\{ \vert a_1' \rangle \langle b'\vert
c' \rangle \otimes \vert d_2' \rangle \pm \vert a_1' \rangle \langle
b' \vert d' \rangle \otimes \langle c_2' \vert \cr &+ \vert c_1'
\rangle \otimes \vert a_2' \rangle \langle b' \vert d' \rangle \pm
\vert d_1' \rangle \otimes \vert a_2' \rangle \langle b' \vert c'
\rangle \bigr\},\cr } \eqno(3.2) $$
where we have assumed the equivalence $\langle b_i' \vert c_i' \rangle \equiv
\langle b'\vert c' \rangle $ for all transformation coefficients.  In
the Abelian case, this result obviously retains its symmetry in the
final state, but it does not in the non-Abelian quaternionic theory.
Symmetrizing $(3.2)$, one finds
$$\eqalign{{\bf M}(a',b') \Psi(c',d') &= {1 \over 2} {1\over \sqrt{2}}
\bigl\{ \vert a_1' \rangle \langle b' \vert c' \rangle \otimes \vert
d_2' \rangle \pm \vert d_1' \rangle \langle b' \vert c'\rangle \otimes \vert
a_2' \rangle \cr &\pm \vert a_1' \rangle \langle b' \vert d' \rangle
\otimes \vert c_2' \rangle + \vert c_1' \rangle \langle b' \vert d'
\rangle \otimes \vert a_2' \rangle \cr &+ \vert c_1' \rangle \otimes
\vert a_2' \rangle \langle b' \vert d' \rangle \pm \vert a_1' \rangle
\otimes \vert c_2' \rangle \langle b' \vert d' \rangle \cr & \pm
\vert d_1' \rangle \otimes \vert a_2' \rangle \langle b' \vert c'
\rangle + \vert a_1' \rangle \otimes \vert d_2' \rangle \langle b'
\vert c' \rangle \bigr\}. \cr} \eqno(3.3) $$
Collecting terms, $(3.3)$ becomes
$$\eqalign{{\bf M}(a',b') \Psi(c',d') &= {1 \over \sqrt{2}}
\bigl\{\vert a_1' \rangle \otimes \vert d_2'\rangle \bullet \langle b'
\vert c' \rangle \pm \vert d_1' \rangle \otimes \vert a_2' \rangle
\bullet \langle b' \vert c' \rangle \cr &\pm \vert a_1' \rangle
\otimes \vert c_2' \rangle \bullet \langle b' \vert d' \rangle + \vert
c_1' \rangle \otimes \vert a_2' \rangle \bullet \langle b' \vert d'
\rangle \bigr\} \cr &= \Psi(a',d') \bullet \langle b' \vert c' \rangle
+ \Psi (c',a') \bullet \langle b' \vert d' \rangle, \cr } \eqno(3.5)$$
where we have defined
$$ \Psi (a',b') \bullet q = {1 \over 2}{1 \over \sqrt{2}} \bigl\{ \vert a'
\rangle q \otimes \vert b' \rangle + \vert a' \rangle \otimes \vert b'
\rangle q \bigr\}, \eqno(3.6)$$
and, similarly, with factor $1/N$ in place of the $1/2$ in $(3.6)$,
for any $N$.  The $\bullet$-product arises from the
requirement of symmetry of the state after the two-body measurement.
We thus find the rule that the generation of a quaternion coefficient
by the action of a filter must be additively distributed, as a
derivation, among the factors of the tensor product in the
representation of an $N$-body state.
\par We remark that the complete Fock space contains the vacuum as
well, and hence $(3.6)$ should be written as
$$ \Psi (a',b') \bullet q = {1 \over 3} \bigl\{ \vert a'
\rangle q \otimes \vert b' \rangle \otimes \vert 0 \rangle +
 \vert a' \rangle \otimes \vert b'
\rangle q \otimes \vert 0 \rangle + \vert a' \rangle \otimes \vert b'
\rangle \otimes \vert 0 \rangle q \bigr\}. \eqno(3.7)$$
\par We now investigate the possibility of closing the algebra of
$${\bf M}(a';q;b')$$ under commutation in a form analogous to that
obtained by Schwinger for the Abelian case, i.e.,
$$ [{\bf M}(a';q;b'),\ {\bf M}(c';p;d')] = {\bf M}(a';q\langle b'
\vert c' \rangle p;d') - {\bf M} (c', p \langle d'\vert a' \rangle q;
b'), \eqno(3.8)$$
where we have suppressed reference to the $\bullet$-product occurring
in sequence at every multiplication.
\par I shall argue the validity of this result, in the limit $N
\rightarrow \infty$, by studying here the structure of the two-body
case, and indicating the form of the general case.
\par Let us apply ${\bf M} (e';p;f')$ to the result $(3.5)$.  One
obtains, in the same way,
$$ \eqalign{{\bf M}(e';p;f')& {\bf M}(a';q;b') \Psi(c',d')\cr
&= {\bf M} (e';p;f') \bigl\{\Psi(a',d') \bullet \langle b' \vert c'
\rangle \bullet q + \Psi(c',a') \bullet \langle b' \vert d' \rangle
\bullet q \bigr\} \cr
&= \Psi(e',d')\bullet \langle b' \vert c' \rangle \bullet q \bullet
\langle f' \vert a' \rangle \bullet p \cr
&+ \Psi (a',e') \bullet \langle b' \vert c' \rangle \bullet q \bullet
\langle f' \vert d' \rangle \bullet p \cr
&+ \Psi(e',a') \bullet \langle b' \vert d' \rangle \bullet q \bullet
\langle f' \vert c' \rangle \bullet p \cr
&+ \Psi (c',e' ) \bullet \langle b' \vert d' \rangle \bullet q \bullet \langle
f'
\vert a' \rangle \bullet p \bigr\}. \cr}
\eqno(3.9) $$
\par In reverse order,
$$ \eqalign{{\bf M}(a';q;b')& {\bf M}(e';p;f') \Psi(c',d')\cr
&= {\bf M} (a';q;b') \bigl\{\Psi(e',d') \bullet \langle f' \vert c'
\rangle \bullet p + \Psi(c',e') \bullet \langle f' \vert d' \rangle
\bullet p \bigr\} \cr
&= \Psi(a',d')\bullet \langle f' \vert c' \rangle \bullet p \bullet
\langle b' \vert e' \rangle \bullet q \cr
&+ \Psi (e',a') \bullet \langle f' \vert c' \rangle \bullet p \bullet
\langle b' \vert d' \rangle \bullet q \cr
&+ \Psi(a',e') \bullet \langle f' \vert d' \rangle \bullet p \bullet
\langle b' \vert c' \rangle \bullet q \cr
&+ \Psi (c',a' ) \bullet \langle f' \vert d' \rangle \bullet p \bullet \langle
b'
\vert e' \rangle \bullet q \bigr\}. \cr}
\eqno(3.10) $$
 \par The difference between these, the commutator, contains the terms
$$ \eqalign{&\Psi(a',e') \bullet \bigl\{\langle b' \vert c' \rangle \bullet q
\bullet \langle f' \vert d' \rangle \bullet p - \langle f' \vert d'
\rangle \bullet p \bullet \langle b' \vert c' \rangle \bullet q
\bigr\}\cr
&+\Psi(e',a') \bullet \bigl\{\langle b' \vert d' \rangle \bullet q
\bullet \langle f' \vert c' \rangle \bullet p - \langle f' \vert c'
\rangle \bullet p \bullet \langle b' \vert d' \rangle \bullet q.
 \cr} \eqno(3.11)$$
Each of these contributions cancels in the Abelian
case.  Due to the lack of commutativity,
cancellation does not, in general, take place in the quaternion case.
However, if we study the effect of the commutator of the generalized
measurement symbols on a many body state ($N \geq 4$), cancellations
 take place among all terms which contain each
factor linearly ( the $\bullet$-product is linearly distributive),
since these factors will occur symmetrically at every position.
The number of these terms is (we include the vacuum factor in the counting)
$${(N+1)!\, \over {(N-3)!\,4!\,}} \sim N^4 $$
as $N \rightarrow \infty$. The remaining terms contain two, three,
 or four (non-commuting) factors at every occupied site, with
frequency of occurrence asymptotically $N^2, \ N^3$ and $N$,
respectively; although these terms do not cancel,
 the four  $\bullet$-products induce a
factor $N^4$ in the denominator, and hence all other combinations
other than the linearly distributed one vanish in the limit (the
linearly distributed terms cancel exactly). The $\bullet$-product of
quaternion coefficients is therefore {\it Abelian} in the limit $N
\rightarrow \infty$.
\par The remaining terms, for which no cancellations occur, close the
algebra in a form precisely analogous to the Abelian case,
$(1.13)$. The corresponding terms in the two-body case, reading from
$(3.9)$ and $(3.10)$, are
$$\eqalign{&\Psi(e',d') \bullet \langle b' \vert c' \rangle \bullet q
\bullet \langle f' \vert a' \rangle \bullet p \cr
&+\Psi(c',e') \bullet \langle b' \vert d' \rangle \bullet q \bullet
\langle f' \vert a' \rangle \bullet p \cr
&-\Psi(a',d') \bullet \langle f' \vert c' \rangle \bullet p\bullet
\langle b' \vert e' \rangle \bullet q \cr
&-\Psi(c',a')\bullet \langle f' \vert d' \rangle \bullet p \bullet
\langle b'\vert e' \rangle \bullet q. \cr} \eqno(3.12)$$
\par In the general case of $N$ particles,
there are $N^4$ non-cancelling contributions to each of these terms,
so dividing by $N^4$ gives a mean value for selection of subsystems
contained, e.g., in $\Psi(d_1', d_2', \dots,d_\ell',\dots, d_N')$
projected to $f'$ or $b'$.  There remain infinite sums, for $N
\rightarrow \infty$, on $i=1,2 \dots N$ containing factors $\langle b'
\vert d_\ell' \rangle, \langle f' \vert d_\ell' \rangle$, and we
assume sufficiently rapid convergence (in the sense of Fourier
transform) as $\ell \rightarrow \infty$.
\par Using the rule of successively applying the $\bullet$-products to the
left, one easily sees that $(3.12)$ arises from the application of
$${\bf M}(e'; p \langle f' \vert a'\rangle q; b') - {\bf M}(a'; q \langle b'
\vert e' \rangle p ; f')$$
to the Fock space vector $\Psi (c',d')$, where we understand that the
quaternionic factors appearing in the arguments must be applied as
$\bullet$-products. The same conclusion follows in the general case.
\par The Schwinger many-body algebra therefore closes
in a form precisely analogous to that of the Abelian case $(1.13)$,
$$\eqalign{&[{\bf M}(e';p;f') {\bf M}(a';q;b') - {\bf M}(a';q;b'){\bf
M}(e';p;f')] \cr
&= {\bf M}(e'; p \langle f'\vert a' \rangle q;b') - {\bf M}(a'; q
\langle b' \vert e' \rangle p;f').\cr } \eqno(3.13)$$
\bigskip
\noindent
{\bf 4. Factorization in terms of quantum fields}
\smallskip
\par In this section, I demonstrate that the annihilation- creation
operators obtained by Razon and Horwitz$^5$, in the limit as the
occupation number $N \rightarrow \infty$, provide an effective
factorization of the quaternionic many-body measurement operators. I
review first the definition of these operators, and show that the
operators $\psi^\dagger(a')\bullet q$ , $\psi (a')\bullet q$,
 where $\psi^\dagger(a')$, $\psi(a')$ are Bose-Einstein or
Fermi-Dirac creation and annihilation operators on the infinite occupation
number
Fock space, are well-defined.  It then follows that
$$ {\bf M}(a';q;b') = \psi(a')^\dagger \bullet q \ \psi(b') \eqno(4.1)$$
provides a representation of the Schwinger algebra $(3.11)$.
\par Razon and Horwitz$^5$ define creation operators on the
$N$-particle Fock space with Bose-Einstein or Fermi-Dirac symmetry
with the property
$$ \psi_N^\dagger(g) \Psi(g_1, \dots, g_N) = \Psi(g_1,\dots,g_N,g),
\eqno(4.2)$$
where I make explicit  the notation
for the (quaternionic) wave functions for the states of the
constituents (the eigenfunction $g^{a'}$ corresponds to $\vert a' \rangle$ in
the
notation used above).  A scalar product between such $\Psi$'s is given
in ref. 5 as a map of pairs $\Psi_1,\, \Psi_2$ into ${\bf H}$ with
positive definite norm (the symplectic tensor product and scalar
products worked out earlier by Biedenharn and Horwitz$^3$ constitute a
special case, but is not a completely quaternion covariant
structure). By means of the Riesz theorem, the corresponding
annihilation operators were found to have the property
$$\eqalign{ &\psi_N(f) \Psi(g_1,\dots,g_N) \cr
&= \sum_{j=1}^N (\pm)^{N-j} \Psi(g_1,\dots,g_{j-1},g_{j+1},\dots,g_N)
\bullet \langle f \vert g_j \rangle_N, \cr}\eqno (4.3)$$
where
$$ \langle f \vert g_j \rangle_N = {1 \over {N+2}} [N\langle f \vert
g_j \rangle + 2 {\rm Re} \langle f \vert g_j \rangle ],\eqno(4.4)$$
and $\langle f \vert g \rangle$ is the quaternionic (one-particle)
scalar product$^{3,4}$. We see from $(4.4)$ the explicit dependence of
these operators on the occupation number of the states.  The
commutation (anticommutation) relations of these operators is given by
$$\eqalign{ [\psi_{N+1} (f) \psi_N^\dagger (g) &\mp {{N+2} \over{N+3}}
\psi_{N-1}(g) \psi_N(f)] \Psi(g_1, \dots,g_N) \cr &=
\Psi(g_1,\dots,g_N) \bullet \langle f \vert g \rangle_{N+1}, \cr &\pm
{1 \over {N+3}} \sum_{j=1}^N (\pm)^{N-j}
\Psi(g_1,\dots,g_{j-1},g_{j+1},\dots,g_N\langle f \vert g_j \rangle ),
\cr} \eqno(4.5)$$
which are deformed$^6$ for finite $N$.
\par It was shown that if the tail of the sequence $\{g_j \},\
j=1,\dots,N$, approaches orthonormality for large $j$, then the norm
of the last term in $(4.5)$ approaches zero as $N\rightarrow \infty$.
It is also true that
$$\langle f \vert g \rangle _N \rightarrow \langle f \vert g \rangle$$
for $N\rightarrow \infty$, and hence the relations $(4.5)$ become, in
this limit,
$$[\psi(f) \psi^\dagger (g) \mp \psi^\dagger(g) \psi(f)] \Psi(g_1,\dots)
= \Psi(g_1, \dots) \bullet \langle f \vert g \rangle . \eqno(4.6)$$
\par The quantum fields, in this limit, become independent of $N$, and
the commutation (anticommutation) relations are no longer deformed.
The $\bullet$-product on the right side of $(4.6)$ corresponds to a
limiting process, and is well-defined (through scalar products) by the
same argument for which the second term on the right hand side of
$(4.5)$ vanishes.
\par Since the state $\Psi$ includes the vacuum as a factor (denoted
by $\otimes 1$ in the algebraic structure of the tensor product
studied in ref. 5),
$$ \Psi(g_1,\dots,g_N) \bullet q = {1 \over {N+1}}\bigl\{ \sum_{j=1}^N
\Psi(g_1, \dots, g_{j-1}, g_jq, g_{j+1}, \dots, g_N) +  \ \
q\Psi(g_1,\dots,g_n) \bigr\}, \eqno(4.7)$$
where the last terms corresponds to multiplication of the vacuum
factor (on the right, if there has been a previous quaternion factor)
by $q$.$^5$  Creating another particle on this state by means of
$\psi^\dagger(f)$, we have
$$\eqalign{\psi^\dagger_N(f) \bigl( \Psi(g,\dots, g_n)\bullet q \bigr)
&= {1 \over {N+1}} \bigl\{ \sum_{j=1}^N
\Psi(g_1,\dots,g_{j-1},g_{j+1},\dots,g_N,f) \cr
 &+ \, q\Psi(g_1,\dots,g_N,f) \bigr\} . \cr} \eqno(4.8)$$
On the other hand,
$$\eqalign {\bigl( \psi^\dagger_N(f) \Psi(g_1,\dots,g_N) \bigr) \bullet
q &= \Psi(g_1,\dots,g_N,f) \bullet q \cr
&= {1 \over {N+2}} \bigl\{ \sum_{j=1}^N
\Psi(g_1,\dots,g_{j-1},g_jq,g_{j+1},\dots,g_N,f) \cr
&+ \,\Psi(g_1,\dots,g_N,fq) +\,q\Psi(g_1,\dots,g_N,f) \bigr\}. \cr}\eqno(4.9)$$
The difference between $(4.8)$ and $(4.9)$ is given by
$$\eqalign{(N+2) \bigl(&\psi_N^\dagger(f) \Psi(g_1,\dots,g_N) \bigr) \bullet q
\cr &-(N+1) \psi_N^\dagger (f) \bigl(\Psi(g_1,\dots,g_N)\bullet
q\bigr) \cr &= \psi_N^\dagger (fq) \Psi(g_1,\dots,g_N). \cr} \eqno(4.10)$$
The right hand side is bounded, as is every term resulting from the
$\bullet$-products, so that for $N \rightarrow \infty$,
$$ \bigl(\psi^\dagger(f) \Psi(g_1,\dots) \bigr) \bullet q = \psi^\dagger
(f) \bigl(\Psi(g_1,\dots)\bullet q \bigr). \eqno(4.11)$$
 It then follows from $(4.10)$, in this limit, that the field
operators have the formal quaternionic linearity property
$$ \psi^\dagger (fq) \Psi(g_1,\dots) = \psi^\dagger(g)
\Psi(g_1,\dots)\bullet q. \eqno(4.12)$$
\par The action of $\psi^\dagger(f)$ is, according to
$(4.11)$, of {\it linear type}$^3$ with respect to
the $\bullet$-product with a quaternion number, and the operator
$$\psi^\dagger(f) \bullet q \, \Psi \equiv \psi^\dagger(f) \Psi
\bullet q $$
is therefore well defined.
\par In a similar way, we may show that
$$ \psi(f)\bigl( \Psi(g_1,\dots)\bullet q \bigr) =
\bigl(\psi(f)\Psi(g_1,\dots) \bigr) \bullet q, \eqno(4.13)$$
and hence the operator $\psi(f)\bullet q$ is also well-defined in the
limiting case of infinite $N$.  To see this, note that typical terms
occurring on the right hand side of $(4.13)$, for finite $N$, are
$$\eqalign{\bigl(&\psi_N(f)\Psi(g_1,\dots,g_N)\bigr)\bullet q \cr
&= \sum_{j=1}^N (\pm)^{N-j}\Psi(g_1,\dots,g_{j-1},g_{j+1},\dots
g_N)\bullet \langle f \vert g_j \rangle \bullet q \cr
&= {1 \over N^2} \sum_{j=1}^N (\pm)^{N-j} \bigl\{\Psi(g_1\langle f
\vert g_j \rangle, g_2,\dots,g_kq,\dots g_{j-1}, g_{j+1},\dots,g_N) \cr
&+ \Psi(g_1, \dots,g_k\langle f \vert g_j \rangle q,\dots, g_{j-1},
g_{j+1},\dots,g_N) \cr
&+\langle f \vert g_j \rangle q \Psi(g_1,\dots, g_{j-1},g_{j+1},\dots,
g_N) \bigr\}. \cr} \eqno(4.14)$$
\par There are
$$ {(N-1)!\, \over 2!\, (N-3)!\, } \sim N^2 $$
terms of the first type, $N$ of the second and one of the third, for
each $j$.  Hence only terms of the first type contribute for $N
\rightarrow \infty$.  For the left side of $(4.13)$, one obtains
$$ \eqalign{ \psi_N(f) &\bigl(\Psi(g_1,\dots,g_N)\bullet q \bigr) \cr
&= {1 \over {N+1}} \bigl\{ \sum_{j=1 \atop k \neq j}^N (\pm)^{N-j}
\Psi(g_1,\dots ,g_kq,\dots, g_{j-1},g_{j+1},\dots,g_N)\bullet \langle f \vert
g_j
\rangle_N \cr
&+\sum_{j=1}^N (\pm)^{N-j} \Psi(g_1,\dots ,g_{j-1},g_{j+1},\dots,g_N)
\bullet \langle f \vert g_jq \rangle_N \cr
&+ q\Psi(g_1,\dots,g_{j-1},g_{j+1},\dots,g_N)\bullet\langle f \vert g_j
\rangle_N \cr
&={1 \over {N(N+1)}} \sum_{j=1 \atop k \neq j}^N (\pm)^{N-j} \bigl\{
\Psi(g_1\langle f
\vert g_j \rangle,\dots g_kq,\dots,g_{j-1},g_{j+1},\dots,g_N) \cr
&+ \cdots +\Psi(g_1,\dots,g_kq\langle f \vert g_j \rangle, \dots,
g_{j-1},g_{j+1},\dots,g_N) \cr
&+q \Psi(g_1,\dots, g_k\langle f \vert g_j \rangle,\dots,
g_{j-1},g_{j+1},\dots,g_N) \cr
&+ \Psi((g_1,\dots, g_k\langle f \vert g_jq \rangle,\dots,
g_{j-1},g_{j+1},,\dots g_N) \cr
&+ \langle f \vert g_j\rangle \Psi(g_1,\dots g_kq,\dots
g_{j-1},g_{j+1},\dots, g_N) \cr
&+{1 \over {N(N+1)}} \sum_{j=1}^N (\pm)^{N-j} \bigl\{
 \langle f \vert g_j q \rangle \Psi(g_1,\dots,g_{j-1},g_{j+1},\dots
g_N) \cr
&+ q\langle f \vert
g_j \rangle \Psi(g_1,\dots, g_{j-1},g_{j+1},\dots, g_N)
 \bigr\}. \cr} \eqno(4.15) $$
Terms of the first type, from distributing $q$ and the annihilation
factor on separate constituents, are ${\rm O} (N^2)$, for each $j$, in
number; terms of the second type, where the distribution of the
annihilation factor coincides with terms with factor $q$, are ${\rm O}
(N)$ for each $j$, as are terms of the third type, from the
distribution of $q$ on the vacuum contribution from the annihilation.
Terms of the fourth type, from the annihilation of $g_jq$ factors are
also ${\rm O}(N)$.  The last two, the vacuum contribution of the
annihilation of $g_jq$ terms, and the product of the two vacuum
factors from $\bullet q$ and the annihilation of $g_j$, are each just
a single term.  In the limit $N \rightarrow \infty $, only terms of
the first type survive, and these are identical to those remaining in
$(4.14)$.
\par We now assert that
$$ {\bf M} (f;q;g) = \psi^\dagger(f) \bullet q \ \psi(g) $$
provides an effective factorization of the many-body measurement
symbols, on states for which $N \rightarrow \infty$.  Consider
$$ \eqalign{ {\bf M}(f';p;g')\,&{\bf M}(f;q;g) \, \Psi(g_1,\dots) \cr
&= \psi^\dagger(f')\bullet p \psi(g') \psi^\dagger (f)\bullet q\, \psi(g)
\Psi(g_1,\dots) \cr
&= \psi^\dagger\bullet p \psi(g')\psi^\dagger(f)\psi(g)
\Psi(g_1,\dots) \bullet q. \cr}\eqno(4.16) $$
Using the identity
$$ \psi(g') \psi^\dagger(f) = \pm\psi^\dagger(f)\psi(g') +
[\psi(g'),\psi^\dagger(f)]_\mp,$$
so that $(4.15)$ becomes, with $(4.6)$,
$$\eqalign{{\bf M}(f';p;g')&{\bf M}(f;q;g) \, \Psi(g_1,\dots) \cr
&= (\pm) \psi^\dagger(f')\bullet p \psi^\dagger(f) \psi(g') \bigl(\psi(g)
\Psi(g_1,\dots) \bigr) \bullet q \cr
&+ \psi^\dagger(f')\bullet p \bigl(\psi(g)\Psi(g_1,\dots) \bigr)\bullet
 q \bullet \langle g' \vert f \rangle \cr
&= (\pm)\psi^\dagger(f')\psi^\dagger(f) \bigl(
\psi(g')\bigl(\psi(g)\Psi(g_1,\dots)\bigr)\bullet q \bigr) \bullet p \cr
 &+ \psi^\dagger (f')\bullet p\bullet \langle g' \vert f \rangle
\bullet q \, \psi(g) \Psi(g_1,\dots). \cr} \eqno(4.17) $$
In reverse order,
$$ \eqalign{{\bf M}(f;q;g)& {\bf M}(f';p;g') \Psi(g_1,\dots)\cr
&= (\pm) \psi^\dagger(f) \psi^\dagger(f') \bigl( \psi(g)
\bigl(\psi(g') \Psi g_1,\dots) \bigr) \bullet p\bigr) \bullet q \cr
&+ \psi^\dagger(f)\bullet q \bullet \langle g \vert f' \rangle \bullet
 p \psi(g') \Psi(g_1,\dots). \cr} \eqno(4.18)$$
The relations $(4.11)$ and $(4.13)$ permit us to eliminate the
association parentheses in the first terms of $(4.17)$ and $(4.18)$,
and the difference is then
$$ \eqalign{{\bf M} (f';p;g')& {\bf M}(f;q;g)- {\bf M} (f;q;g) {\bf M}
(f';p;g')] \Psi(g_1,\dots) \cr
&= (\pm)\psi^\dagger (f') \psi^\dagger(f) \psi(g') \psi(g)
\Psi(g_1,\dots)\bullet (q \bullet p - p\bullet q) \cr
&+\bigl\{\psi^\dagger(f')\bullet p \bullet \langle g' \vert f \rangle
\bullet q \psi(g) \cr
&- \psi^\dagger(f) \bullet q \bullet \langle g \vert f' \rangle
\bullet p \psi(g') \bigr\} \Psi(g_1,\dots). \cr} \eqno(4.19) $$
The first term of $(4.19)$, in the distributive product commutator,
contains, as in $(3.11)$, cancellations of the ${\rm O}(1)$ terms, and
hence vanishes in the $N \rightarrow \infty$ limit.  The second term
of $(4.19)$ reproduces $(3.12)$.
\bigskip
\noindent
{\bf 5. Discussion}
\smallskip
\par The Schwinger algebra, based on fundamental notions of
measurements and the kinematical independence of elements of the
quantum ensemble, can be realized in terms of quaternionic quantum
theory as well as in the framework of the usual complex Hilbert
space.  Extending the algebra to the many-body theory, Schwinger found
that the Bose-Einstein and Fermi-Dirac quantum fields of the complex
theory provide a representation of the commutation relations of the
many-body algebra independently of the class of states of the theory.
In the quaternionic case, I find that the many-body algebra cannot be
closed consistently in the finite particle sector of the Fock space.
If all physical states of the theory contain an infinite number of elementary
excitations, i.e., in the infinite particle sector of the Fock space
(the Haag theorem$^{12}$ asserts that there is no unitary
transformation that connects the finite and infinite sectors), the
algebra closes in Schwinger's form. In this sector, the distributive
$\bullet$-product of quaternion coefficients constitutes an
associative, Abelian algebra. A quaternion factor $q$
occurring as a right multiplier of any constituent state in the tensor
product may be extracted as a $\bullet$-product to the right of the
Fock space vector[$(4.12)$]; since such multipliers are Abelian (in the
$N \rightarrow \infty $ limit), the result of
extracting several such factors is independent of the order or their
source.  The $\bullet$-product of quaternion factors (in the infinite
sector) is not
homomorphic to the quaternion algebra ${\bf H}$, or to any finite
algebra; it does not
 close.
\par  The
annihilation-creation operators of the quaternionic Fock space defined
by Razon and Horwitz$^5$ provide an effective representation for the
factorization of this algebra in the limit $N \rightarrow \infty$.  It
is only in this limit that the commutation (anticommutation) relations
of these operators are not deformed, and coincide formally with the
commutation (anticommutation) relations of complex quantum field
theory. The numerical coefficient arising from the commutation
(anticommutation) of operators associated with different one particle states
appears as a $\bullet$-product on vectors of the Fock space as well.
\par  I have demonstrated that in the infinite sector, the
distributed right multipliers may be considered as part of the action
of the field operators themselves, which then form, in this way, an
algebra with many of the properties of the usual complex field. Since
the commutation (anticommutation) relations of the fields are of the
same structure as in the usual complex theory, the treatment of
dynamical problems, such as the construction of the $S$-matrix by
diagrammatic techniques becomes accessible (with some suitable
definition of the ground state, as in condensed matter theory).
\par In the case of finite $N$, the commutation (anticommutation)
relations are deformed, and the Schwinger algebra is not satisfied for
the many-body theory.  The physical meaning of this sub-asymptotic
regime, and  the relation of the structure discussed here to
the work of Adler and Millard$^{11,13}$, remain to be investigated.
\bigskip
\noindent
{\it Acknowledgements}
\par This work was initiated, and some of the preliminary results
obtained, at the Institute for Advanced Study,
Princeton, New Jersey.  I would like to thank S.L. Adler for his hospitality
there, and for many discussions on this and related
topics. I am also grateful to A. Razon for the many discussions
during our work on the construction of the tensor product that
 led to the ideas and techniques that I applied here.
\par
\bigskip
{\bf References}
\frenchspacing
\smallskip
\item{1.} J.S. Schwinger, {\it Quantum Kinematics and Dynamics,\/}
W.A. Benjamin, New York (1970), and lectures at Harvard University (1954).
\item{2.} D. Finkelstein, J.M. Jauch, S. Schiminowich and D. Speiser,
Jour. Math. Phys. {\bf 3}, 207 (1962); {\bf 4}, 788 (1963).
\item{3.} L.P. Horwitz and L.C. Biedenharn, Ann. Phys. {\bf 157}, 432
(1984).
\item{4.} S.L. Adler {\it Quaternionic Quantum Mechanics and Quantum
Fields,\/} Oxford University Press, New York and Oxford (1995).
\item{5.} A. Razon and L.P. Horwitz, Acta Applicandae Mathematicae
{\bf 24}, 141 (1991); {\bf 24}, 179 (1991).
\item{6.} For example, L.C. Biedenharn and M.A. Lohe, {\it Quantum Group
Symmetry
and q-Tensor Analysis,\/} World Scientific (1995); S. Shnider and
S. Sternberg, {\it Quantum Groups,\/} International Press, Hong Kong
(1993);
C. Kassel, {\it Quantum Groups,\/} Springer Verlag, New York (1995).
\item{7.} A. Connes, {\it Noncommutative Geometry,\/}Academic Press,
San Diego (1994).
\item{8.} J. Madore, {\it An Introduction to Noncommutative
Differential Geometry and its Physical Applications,\/} London
Math. Soc. Lecture Note Series 206, Cambridge Univ. Press (1993).
\item{9.} C. Piron, {\it Foundations of Quantum
Physics,\/}W.A. Benjamin, Reading, Mass. (1976).
\item{10.} S.L. Adler, Nuc. Phys. B {\bf 415}, 195 (1994).
\item{11.} S.L. Adler and A. Millard, Nuc. Phys. B {\bf 473}, 199 (1996).
\item{12.} S.L. Adler and L.P. Horwitz, Jour. Math. Phys. {\bf 37},
5429 (1996) (hep-th/9606023);
S.L. Adler and L.P. Horwitz, Proc. of Workshop on
``Algebraic Approaches to Quantum Dynamics,'' May 7-12, Fields
Institute for Research in Mathematical Sciences, Toronto, Ontario, CA,
hep-th/9606019; S.L. Adler and L.P. Horwitz, Proc. of Quantum
Structures '96 Berlin, July 29-Aug.3,1996, hep-th/9610153.
\item{13.} R. Haag, Dan. Mat. Fys. Medd. {\bf 29}, 12 (1955); D.Hall and
A.S. Wightmann, Mat. Fys. Medd. Dan. Vid. Selsk, {\bf 31}, 5 (1957).
See also, R.F. Streater and A.S. Wightmann, {\it PCT, Spin and
Statistics and All That,\/} W.A. Benjamin, New York (1964).

\vfil
\eject
\bye
\end